# CellSense: An Accurate Energy-Efficient GSM Positioning System

Mohamed Ibrahim, *Student Member, IEEE,* and Moustafa Youssef, *Senior Member, IEEE*

*Abstract*—Context-aware applications have been gaining huge interest in the last few years. With cell phones becoming ubiquitous computing devices, cell phone localization has become an important research problem. In this paper, we present CellSense, a probabilistic RSSI-based fingerprinting location determination system for GSM phones. We discuss the challenges of implementing a probabilistic fingerprinting localization technique in GSM networks and present the details of the CellSense system and how it addresses these challenges. We then extend the proposed system using a hybrid technique that combines probabilistic and deterministic estimation to achieve both high accuracy and low computational overhead. Moreover, the accuracy of the hybrid technique is robust to changes in its parameter values. To evaluate our proposed system, we implemented CellSense on Android-based phones. Results from two different testbeds, representing urban and rural environments, for three different cellular providers show that CellSense provides at least **108.57%** enhancement in accuracy in rural areas and at least **89.03%** in urban areas compared to the current state of the art RSSI-based GSM localization systems. In additional, the proposed hybrid technique provides more than 6 times and 5.4 times reduction in computational requirements compared to the state of the art RSSI-based GSM localization systems for the rural and urban testbeds respectively. We also evaluate the effect of changing the different system parameters on the accuracy-complexity tradeoff and how the cell towers density and fingerprint density affect the system performance.

## I. INTRODUCTION

As cell phones become more ubiquitous in our daily lives, the need for context-aware applications increases. One of the main context information is location which enables a wide set of cell phone applications including navigation, location-aware social networking, and security applications. Although GPS [2] is considered one of the most well known localization techniques, it is not available in many cell phones, requires direct line of sight to the satellites, and consumes a lot of energy. Therefore, research for other techniques for obtaining cell phones location has gained momentum fueled by both the users' need for location-aware applications and government requirements, e.g. FCC [3]. City-wide WiFi-based localization for cellular phones has been investigated in [4], [5] and commercial products are currently available [6]. However, WiFi chips, similar to GPS, are not available in many cell phones and not all cities in the world contain sufficient WiFi coverage to obtain ubiquitous localization. Similarly, using augmented sensors in the cell phones, e.g. accelerometers and compasses, for localization have been proposed in [7]–[9]. However, these sensors are still not widely used in many phones. On the other hand, GSM-based localization, by definition, is available on all GSM-based cell phones, which presents 80-85% of today's cell phones [10], works all over the world, and consumes minimal energy in addition to the standard cell phone operation. Many research work have addressed the problem of GSM localization [3], [5], [11], [12], including time-based systems, angle-of-arrival based systems, and received signal strength indicator (RSSI) based systems. Only recently, with the advances in cell phones, GSM-based localization systems have been implemented [5], [11], [12]. These systems are mainly RSSI-based as RSSI information is easily available to the user's applications. Since RSSI is a complex function of distance, due to the noisy wireless channel, RSSI-based systems usually require building an RF fingerprint of the area of interest [5], [11], [12]. A fingerprint stores information about the RSSI received from different base stations at different locations in the area of interest. This is usually constructed once in an offline phase. During the tracking phase, the received RSSI at an unknown location is compared to the RSSI signatures in the fingerprint and the closest location in the fingerprint is returned as the estimated location. Constructing the fingerprint is a time consuming process. However, this is typically done in a process called war driving, where cars scan the streets of a city to map it. Current commercial systems, such as Skyhook, Google's MyLocation and StreView services already perform scanning for other purposes. Therefore, constructing the fingerprint for GSM localization can be *piggybacked on these systems* **without extra overhead**.

In this paper, we propose *CellSense*, a **probabilistic** fingerprinting based technique for GSM localization. Unlike the current fingerprinting techniques for GSM phones that use a deterministic approach for estimating the location of cell phones [11], [12], the *CellSense* probabilistic technique provides more accurate localization. However, constructing a probabilistic fingerprint is challenging, as This work is supported in part by a Google Research Award.
M. Ibrahim is with the Wireless Intel. Net. Center (WINC), Nile University, Smart Village, Egypt e-mail: m.ibrahim@nileu.edu.eg.
M. Youssef is with the Dep. of Comp. Sc. and Eng., Egypt-Japan Univ. of Sc. & Tech. (E-JUST), Alexandria, Egypt e-mail: moustafa.youssef@ejust.edu.eg.
An earlier version of this paper has appeared in the proceedings of the IEEE Global Communications Conference (GlobeCom) 2010 [1].

we need to stand at each fingerprint location for a certain amount of time to construct the signal strength histogram. This adds significantly to the overhead of the fingerprint construction process. *CellSense* addresses this challenge by using gridding, where the area of interest is divided into a grid and the histogram is constructed for each grid cell. This, not only removes the extra overhead of standing at each location for a certain time, but also helps in increasing the scalability of the technique as the fingerprint size can be reduced arbitrarily by increasing the grid cell length. To further reduce the computational overhead of *CellSense*, we propose a hybrid technique *CellSense-Hybrid* that combines a probabilistic estimation phase with a deterministic refinement phase. The *CellSense-Hybrid* technique has also the added advantage of its accuracy being robust to changes in its parameter values.

In order to evaluate *CellSense*, we implement it on Android-enabled cell phones and compare its performance to other deterministic fingerprinting techniques, model based techniques, and Google's MyLocation service under two different testbeds representing rural and urban environments for three different cellular providers. We also study the effect of the different parameters on the performance of *CellSense*. Our results show that *CellSense* outperforms other systems, with at least 108.57% and 89.03% enhancement in accuracy for the urban and rural testbeds respectively. In addition, it has significant savings in terms of energy consumption with 5 to 6 times saving in running time. Moreover, the *CellSense-Hybrid* technique accuracy is robust to the changes in parameter values.

To summarize, the contribution of this paper is three-fold:

1) We introduce the *CellSense* probabilistic GSM localization system. *CellSense* provides high localization accuracy and depends on a novel gridding technique to reduce the fingerprint construction overhead.
2) We further extend the *CellSense* technique through a hybrid technique that adds a deterministic refitment phase to the basic *CellSense* technique. The accuracy of the *CellSense-Hybrid* technique is robust to changes in its parameter values. Therefore, the *CellSense-Hybrid* technique parameters can be selected to achieve a low computational overhead while maintaining the same accuracy.
3) We thoroughly evaluate the performance of the *CellSense* and *CellSense-Hybrid* techniques, both through analysis and under two different testbeds, and show their significant advantage compared to other state-of-the-art GSM localization systems.

The rest of the paper is organized as follows: In Section II we discuss relevant related work. In Section III, we present our *CellSense* system. Section IV presents the performance evaluation of our system. Finally, Section V concludes the paper and gives directions for future work.

,

## II. Related Work

In this section, we discuss the different techniques for cell phone localization and how they differ from the proposed work. We categorize these techniques as: time-based, angle of arrival based, cell-id based, city-wide WiFi localization, augmented sensors based, and signal strength based.

### A. Time-of-Arrival based Localization

In time-of-arrival (ToA) based systems, the cell phone estimates its distance to a reference point based on the time a signal takes to travel from the reference point to it. Similarly, time difference of arrival (TDOA) based systems use the principle that the emitter location can be estimated by the intersection of the hyperbolae of constant differential time of arrival of the signal at two or more pairs of base stations [3].

The most well known localization technique, the GPS [2], can be categorized as a time-of-arrival based system. Time based systems require special hardware and therefore are usually deployed on high-end phones. In addition, GPS suffers from two other main problems: availability and power consumption; It requires line-of-sight to the satellites; therefore it does not work indoors and it consumes a lot of power of the energy-limited cell phones.

### B. Angle-of-Arrival based Systems

Angle-of-Arrival (AOA) based systems use triangulation based on the estimated AOA of a signal at two or more base stations to estimate the location of the desired transmitter [3], [13]–[16]. Antenna arrays are usually used to estimate the angle of arrival. Similar to TOA based systems, AOA based systems require specialized hardware, which makes them less attractive for a large deployment on cell phones.

### C. Cell-ID based Techniques

Cell-ID based techniques, e.g. Google's MyLocation [17], do not use RSSI explicitly, but rather estimate the cell phone location as the location of the cell tower the phone is currently associated with. This is usually the cell tower with the strongest RSSI. Such techniques require a database of cell towers locations and provide an efficient, though coarse grained, localization method.

### D. City-wide WiFi-based localization

City-wide WiFi-based localization has been proposed in [4], [5] and commercial products are currently available, e.g. [6]. However, WiFi chips, similar to GPS, are not available in the majority of cell phones and not all cities in the world contain sufficient WiFi coverage to obtain ubiquitous localization.

### E. *Augmented Sensors-based localization*

Using augmented sensors in the cell phones, e.g. accelerometers and compasses, for localization have been proposed in [7]–[9], [18]. For example, in [18] the authors use the accelerometer and compass as an energy-efficient way for estimating the phone displacement and direction. Due to the accumulation of error, they synchronize with the GPS as needed.

The main issue with augmented sensors-based localization systems is that these sensors are still not widely used in cell phones.

### F. *RSSI-based Systems*

Recently, RSSI-based systems have been introduced and implemented for cell phone localization. Since RSSI information is readily available to the user's applications on almost all GSM phones, such systems have the potential of localizing 80-85% of today's cell phones [10], work all over the world, and consume minimal energy in addition to the standard cell phone operation.

However, since RSSI is a complex function of distance, RSSI-based systems usually require building an RF fingerprint of the area of interest [5], [11], [12]. A fingerprint stores information about the RSSI received from different base stations at different locations in the area of interest. This is usually constructed once in an offline phase. During the tracking phase, the received RSSI at an unknown location is compared to the RSSI signatures in the fingerprint and the closest location in the fingerprint is returned as the estimated location. Constructing the fingerprint is a time consuming process. However, this is typically done in a process called war driving, where cars drive the area of interest continuously scanning for cell towers and recording the cell tower ID, RSSI, and GPS location. Current commercial systems, such as Skyhook, Google's MyLocation and StreeView services already perform scanning for other purposes. Therefore, constructing the fingerprint for GSM localization can be *piggybacked on these systems* **without extra overhead**.

In the rest of this section, we summarize the current work in fingerprint-based RSSI localization systems for GSM phones, which is the closest to the proposed work.

*1) Deterministic Fingerprinting Techniques:* Current fingerprinting techniques for GSM localization use only deterministic techniques [11], [12]. For example, each location in the fingerprint of [11] stores a vector representing the RSSI value from each cell tower heard at this location. During the tracking phase, the K-Nearest Neighbors (KNN) classification algorithm is used, where the RSSI vector at an unknown location is compared to the vectors stored in the fingerprint and the K-closest fingerprint locations, in terms of Euclidian distance in the RSSI space, to the unknown vector are averaged as the estimated location. Deterministic fingerprinting techniques require searching a larger database than cell-ID based techniques but provide higher accuracy. Note that the overhead of constructing the fingerprint is the same as constructing the cell-ID database as both require war driving.

*2) Modeling-based Techniques:* Modeling-based techniques try to capture the relation between signal strength and distance using a model. For example, the work in [11] uses a Gaussian process to capture this relation assuming that the received signal strength $y_i$ at location $x_i$ is $y_i = f(x_i) + \epsilon_i$ Where $\epsilon_i$ is zero mean, additive Gaussian noise with known variance $\sigma_n^2$.

A Gaussian process (GP) estimates posterior distributions over functions $f$ from a training data $D$ (fingerprint). These distributions are represented non-parametrically, in terms of the training points. A key idea underlying GP's is the requirement that the function values at different points are correlated, where the covariance between two function values, $f(x_p)$ and $f(x_q)$, depends on the input locations, $x_p$ and $x_q$. This dependency can be specified via an arbitrary covariance function, or kernel, $k(x_p, x_q)$. The most widely used kernel function is the squared exponential, or Gaussian, Kernel: $k(x_p, x_q) = \sigma_f^2 exp(\frac{-1}{2l^2}|x_p - x_q|^2)$, where $l$ is a length scale that determines how strongly the correlation between points drops off.

Building a GP estimator still requires constructing a fingerprint, though a less sparse one. This fingerprint is used to estimate the model parameters ($l$, $\sigma_n^2$, and $\sigma_f^2$) and to compute $f(x_*)$ for any location $x_*$.

This reduces the size of the fingerprint and provides a way for extending a sparse fingerprint to a more dense one as it gives the fingerprint values at any arbitrary location based on the assumed model. However, this comes at the cost of substantial increase in computational requirements, as we quantify in Section IV, and there is no actual saving of fingerprinting overhead as war driving has to be done to collect the training samples ($D$) anyway. Moreover, the assumed model may not fit the real environment, thus reducing the accuracy of the returned location.

### G. *Summary*

Compared to TOA, AOA, city-wide WiFi, and augmented sensors based systems, our proposed system, *CellSense*, does not require any specialized hardware and is more ubiqtious, in terms of the number of cell phones it runs on and the coverage area.

Compared to the cell-ID based systems and the current fingerprinting techniques, our technique is a **probabilistic** one. Using a probabilistic approach should enhance the accuracy of localization compared to a deterministic approach. However it comes with its own challenges, such as constructing the RSSI probability distribution with minimal overhead. Our proposed technique addresses these challenges and provides accuracy better than all of the current techniques with minimal computational requirements as we quantify in Section IV.

## III. THE CELLSENSE SYSTEM

In this section, we describe our *CellSense* system for GSM phones localization. We start by an overview of the system followed by the details of the offline training and online tracking phases. Finally, we propose a hybrid approach that combines the basic *CellSense* and a deterministic approach to achieve both accurate localization and low computational overhead.

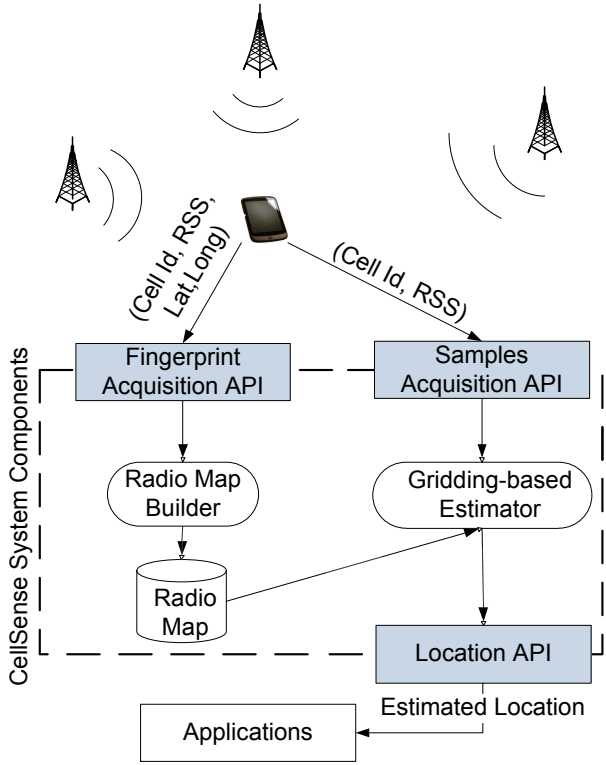

Figure 1. *CellSense* components: the arrows show information flow in the system.

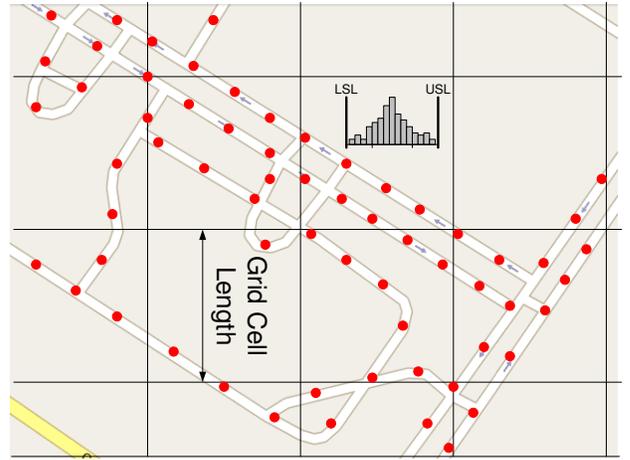

Figure 2. *CellSense* approach for fingerprint construction. The area of interest is divided into grids and the histogram is constructed using the fingerprint locations inside the grid cell. No extra overhead is required for fingerprint construction. The grid cell length parameter can be used to tradeoff accuracy and scalability.

### A. Overview

Figure 1 shows our *CellSense* system architecture. *CellSense* works in two phases: an offline fingerprint construction phase and online tracking phase. During the offline phase, a probabilistic fingerprint is constructed, where the RSSI histogram for each cell tower at given locations in the area of interest is estimated. This is performed in the *Radio Map Builder* module.

During the online tracking phase, the location estimation module uses the fingerprint to calculate the most probable fingerprint location the user may be standing at.

The RSSI samples are collected with the *Fingerprint Acquisition API* that interacts with the phone GSM modem to obtain RSSI information from up to seven neighboring cell towers as indicated by the GSM standard. Finally, the *Location API* is used by the user's applications to query the current estimated user's location.

### B. Mathematical Model

Without loss of generality, let $\mathbb{L}$ be a two dimensional physical space. Let $q$ represent the total number of cell towers in the system. We denote the $q$-dimensional signal strength space as $\mathbb{Q}$. Each element in this space is a $q$-dimensional vector whose entries represent the RSSI readings from a different cell tower. We refer to this vector as $s$. We also assume that the samples from different towers are independent. Therefore, the problem becomes, given an RSSI vector $s = (s_1, ..., s_q)$, we want to find the location $l \in \mathbb{L}$ that maximizes the probability $P(l|s)$.

### C. Offline Phase

The purpose of this phase is to construct the signal strength histogram for the RSSI received from each cell tower at each location in the fingerprint. Typically, this requires the user to stand at each location in the fingerprint for a certain period of time to collect enough samples to construct the RSSI histogram. This will increase the fingerprint construction overhead significantly, as the wardriving car has to stop at each location in the fingerprint for a certain time.

To avoid this overhead, we use a gridding approach, where the war-driving process is performed normally and the area of interest is divided into cells. The histogram is then constructed for each cell tower in a given cell using all fingerprint points inside the cell, rather than for each individual fingerprint point (Figure 2). Note that this gridding approach reduces the resolution of the fingerprint from individual points to cells with a certain size. The center of mass of all fingerprint points inside a grid cell is used to represent the cell[1]. Figure 3 shows the histograms for a certain cell tower in three adjacent cells. The figure shows that the shape of the histogram changes over the different grid cells and hence could be used to distinguish between them.

The gridding approach not only removes the extra overhead of war-driving, but also increases the scalability of *CellSense* as the fingerprint size can be arbitrarily reduced by increasing the cell size. We quantify the effect of the grid cell length parameter on performance in Section IV.

---

[1]We use the term "fingerprint point" to refer to an individual point collected by the wardriving car and use the term "fingerprint cell" to denote the fingerprint collected using all points inside a given cell.

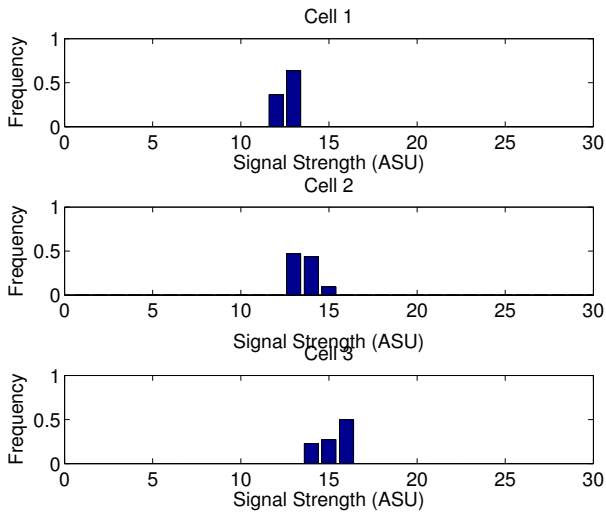

Figure 3. An example of the histograms from three adjacent cells (grid length= 70m) from a certain cell tower. The Active Set Update (ASU) is an integer value returned by the phone API ($dBm = 2.ASU - 113$).

## D. Online Phase

During the online phase, the user is standing at an unknown location $l$ receiving a signal strength vector $s = (s_1, ..., s_q)$, containing one entry for each cell tower. We want to find the location in the fingerprint ($l \in \mathbb{L}$) that has the maximum probability given the received signal strength vector $s$. That is, we want to find

$$argmax_l[P(l|s)] \quad (1)$$

Using Bayes' theorem and assuming that all locations are equally probable[2], this can be written as:

$$argmax_l[P(l|s)] = argmax_l[P(s|l)] \quad (2)$$

$P(s|l)$ can be calculated using the histograms constructed during the offline phase as:

$$P(s|l) = \prod_{i=1}^{q} P(s_i|l) \quad (3)$$

The above equation considers only one sample from each stream for a location estimate. In general, a number of successive samples, $N_s$, from each stream can be used to improve performance.

In this case, $P(s|l)$ can then be expressed as follows:

$$P(s|l) = \prod_{i=1}^{q} \prod_{j=1}^{N} P(s_{i,j}|l) \quad (4)$$

Where $s_{i,j}$ represents the $j^{th}$ sample from the $i^{th}$ stream. Thus, given the signal strength vector $s$, the discrete space estimator applies Equation 4 to calculate $P(s|l)$ for each location $l$ and returns the location that has the maximum probability.

---

[2]If the probability of being at each location is known, this can be used in the equation as is.

Similarly, instead of returning just the most probable location, a weighted average of the $K$ most probable fingerprint cells, weighted by the probability of each location, can be used to obtain a better estimate of location. We study the effect of the parameter $K$ on performance in Section IV.

## E. The CellSense-Hybrid Technique

For the described *CellSense* technique, the grid cell length parameter allows us to trade accuracy and computational complexity: Larger cells lead to lower accuracy but they reduce the computational complexity due to the reduced number of cells. The *CellSense-Hybrid* technique targets maintaining the accuracy at lower grid sizes while reducing the computational requirements. To achieve both accuracy and low complexity, the *CellSense-Hybrid* technique runs in two phases: Rough estimation phase and refinement phase.

1) In the first phase (rough estimation phase), it uses the standard probabilistic fingerprint estimation technique to obtain the most probable cell a user may be located in. However, instead of returning the center of mass of the fingerprint points inside this cell as the estimated location as in the standard *CellSense*, it refines this estimate in the second step.
2) In the second phase (estimation refinement phase), a K-nearest neighbor approach is used to estimate the closest fingerprint point, in the signal strength space, to the current user location inside the cell estimated in phase one. Note that since the histograms are constructed for an entire cell, we do not use a probabilistic technique in the second phase.

To achieve a low computational cost at low values of the grid cell length parameter, the *CellSense-Hybrid* technique uses only one sample to estimate the most probable cell, rather than $N_s$ samples, in its first phase. The refinement phase allows it to compensate for the lost accuracy. Note that the *CellSense-Hybrid* technique does not have an advantage, in terms of computational complexity, for higher grid cell lengthes as the number of fingerprint points involved in the second phase will dominate the computational cost.

In summary, the low computational requirement of the *CellSense-Hybrid* technique is achieved by using a fewer number of samples in the estimation process as compared to the *CellSense*. To compensate for the reduced accuracy, *CellSense-Hybrid* uses an estimation refinement phase. This allows *CellSense-Hybrid* to achieve both high accuracy and low computational requirements for low values of the grid cell length parameter. We quantify the performance of the hybrid technique in Section IV.

## IV. PERFORMANCE EVALUATION

In this section, we study the effect of different parameters on *CellSense* and compare its performance to other RSSI-based GSM localization systems in terms of localization accuracy and running time. For the running time estimation, all techniques have been implemented on a Dell Inspiron 6400 with a 1.83GHz Intel Core 2 processor running Windows XP.

## A. Data Collection

We collected data for two different testbeds. The first testbed covers the Smart Village in Cairo, Egypt which represents a typical rural area. The second testbed covers a 5.45 Km$^2$ in Alexandria, Egypt representing a typical urban area. Data was collected using T-Mobile G1 phones which have a GPS receiver (used as the ground truth for location) and running the Android 1.6 operating system. The experiment was performed using three phones, each with a SIM card for a different cellular provider in Egypt.

We implemented the scanning program using the Android SDK. The program records the (cell-ID, signal strength, GPS location, timestamp) for the cell tower the mobile is connected to as well as the other six neighboring cell towers information as dedicated by the GSM specifications. The scanning rate was set to one per second. Two independent data sets were collected for each testbed: one for training and the other for testing. Table I summarizes the two testbeds.

The calibration process took on average 22.34 minutes for the rural area and 48.48 minutes for the urban area. The war-driving process involved visiting each point only once. Our experience show that visiting the same point more than one time does not lead to enhancement in accuracy.

## B. Effect of Changing Parameters

In this section, we explore the results of changing the different parameters on the performance of *CellSense*, mainly: grid cell length, number of samples used in estimation ($N_s$) and the number of most probable locations averaged to obtain the final location ($K$). We also study the effect of changing the network provider, cell towers density, and the effect of using a sparse radio map. Table II summarizes the parameters and their default values, which are the values that achieve the best performance.

*1) Effect of grid cell length:* Figure 4 shows the effect of changing the grid cell length on the median localization error. Each cell is a square with size as indicated on the x-axis. The figure shows that as the cell size increases, the accuracy decreases. This is because as the grid cell length increases the points inside a cell become further away from its centroid, increasing the estimation error.

The figure also shows that a grid cell length up to 200 $m^2$ gives comparable accuracy to very small cell sizes for both testbeds. This indicates that *CellSense* can lead to good scalability with minimal reduction in accuracy. Moreover, the figure shows that the accuracy in urban areas is better than the accuracy in rural areas for grid cell length up to 450m due to the increased cell tower density. Increasing the grid cell length beyond this value leads to a significant drop in performance for the urban testbed. We believe this is due to the fact that cell towers are configured to have a smaller range in urban areas. Increasing the cell length size beyond a certain value makes some cell towers not cover an entire cell, increasing the ambiguity between cells and reducing accuracy.

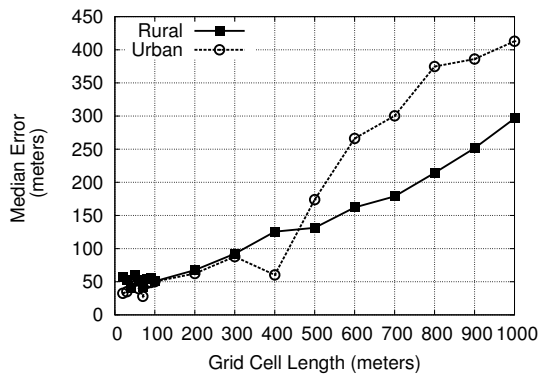

Figure 4. Effect of changing the grid cell length on *CellSense*'s median error.

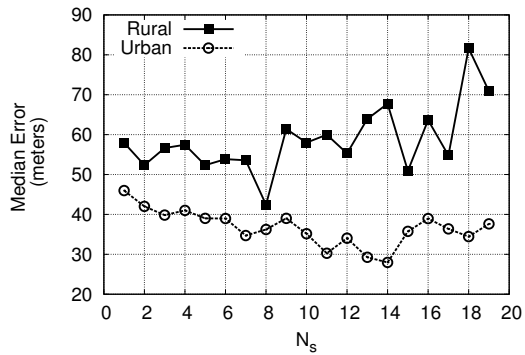

Figure 5. Effect of changing the number of samples ($N_s$) on *CellSense*'s median error.

*2) Effect of the number of samples used ($N_s$):* Figure 5 shows the effect of changing the number of samples used in estimation ($N_s$) on the median localization error. The figure shows that as the number of samples used in estimation increases, the accuracy increases until it reaches an optimal value ($N = 8$ and $N = 14$ for the rural and urban testbeds respectively) and then decreases. This is due to two opposing factors: (1) As we increase the number of samples, we have more information to estimate the user location and hence we should get better accuracy. (2) However, as we increase the number of samples, the time to collect these samples increases which leads to crossing the boundary of one cell when using a large number of samples. This has a negative effect on accuracy.

The optimal point in rural areas occurs at lower $N_s$ compared to the urban areas due to the fact that the user speed is higher in rural areas than in urban areas.

*3) Effect of the number of averaged fingerprint locations ($K$):* Figure 6 shows the effect of changing the number of the most probable locations averaged ($K$) on the median localization error. The figure shows that, in general, the performance enhances as $K$ increases until it saturates. This also highlights that the most probable location estimate has a good accuracy.

| Testbed | Area covered (Km²) | Trace length (Km) | Network provider | Average calibration time(min.) | Avg. fgrprt dens/ cell | Total num. of cell towers | Training set size | Test set size | Avg. num. towers / loc. | Avg. num. of towers / Km² | Avg. num. of towers /Km |
|---|---|---|---|---|---|---|---|---|---|---|---|
| One (Rural) | 1.958 | 13.64 | Provider 1 | 26.65 | 10.89 | 51 | 1599 | 573 | 5.63 | 26.05 | 3.05 |
| | | | Provider 2 | 17.88 | 9.17 | 11 | 1073 | 594 | 1.00 | 5.62 | 1.04 |
| | | | Provider 3 | 22.51 | 12.79 | 59 | 1351 | 592 | 4.83 | 30.13 | 4.33 |
| Two (Urban) | 5.450 | 18.27 | Provider 1 | 51.5 | 11.22 | 137 | 3090 | 1239 | 5.35 | 25.13 | 6.97 |
| | | | Provider 2 | 48.9 | 11.07 | 121 | 2934 | 1560 | 6.01 | 22.20 | 6.56 |
| | | | Provider 3 | 45.06 | 12.12 | 155 | 2704 | 1564 | 5.32 | 28.44 | 9.28 |

Table I
COMPARISON BETWEEN THE TWO TESTBEDS. THE TRAINING SET SIZE REFERS TO THE NUMBER OF SAMPLES COLLECTED BY THE WAR DRIVING PROCESS. THE AVERAGE FINGERPRINT DENSITY IS THE AVERAGE NUMBER OF FINGERPRINT POINTS INSIDE A CELL FOR GRID CELL LENGTH= 70M.

| Parameter | CellSense (Best accuracy) | CellSense-Hybrid (**Best timing**) | Deterministic (Best accuracy) | Gaussian Processes (Best accuracy) |
|---|---|---|---|---|
| Rural testbed | Grid size=70, $N = 14$, $K = 2$ | Grid size=70, $K = 1$ | Grid size=70, $K = 8$ | $N_p = 1019$ |
| Urban testbed | Grid size=70, $N = 8$, $K = 2$ | Grid size=70, $K = 1$ | Grid size=90, $K = 6$ | $N_p = 573$ |

Table II
DEFAULT VALUES FOR THE PARAMETERS. THESE VALUES ACHIEVES THE BEST PERFORMANCE.

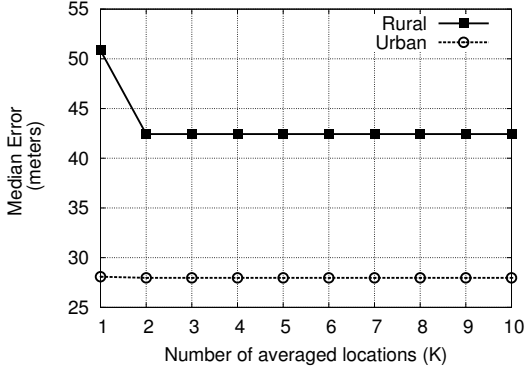

Figure 6. Effect of changing the number of most probable locations averaged ($K$) on *CellSense*'s median error.

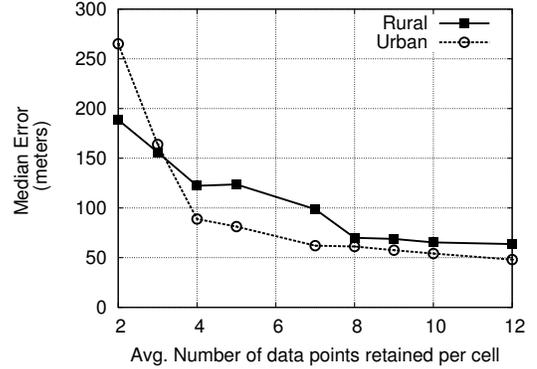

Figure 8. Effect of reducing the average number of data point per cell *CellSense*'s median error.

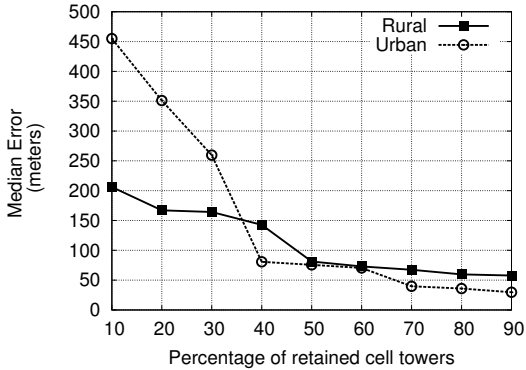

Figure 7. Effect of changing cell towers density on *CellSense*'s median error.

*4) Effect of changing the cell towers density:* Figure 7 shows the effect of changing the cell towers density on the median localization error. This was achieved by dropping a certain percentage of the cell towers as indicated in the figure. The figure shows that as the cell towers density increases, the accuracy increases.

*5) Effect of decreasing the radio map density:* Figure 8 shows the effect of decreasing the fingerprint density on the median localization error. The figure shows that as the percentage of retained samples increases, the accuracy increases. The figure also shows that collecting only 8 points per cell is enough to obtain good accuracy for both testbeds. In addition, the effect of reducing the fingerprint density is less than the effect of reducing the cell tower density.

*6) Effect of using different network providers:* Figure 9 shows the effect of using different network providers in rural and urban areas. The figure shows that the accuracy of the provider is proportional to its cell tower density reported in Table I. The noticeable difference between Provider 2 and the other two providers in the rural testbed is due to its significantly lower cell tower density per location (1.00 as compared to 4.83 and 5.63). In addition,

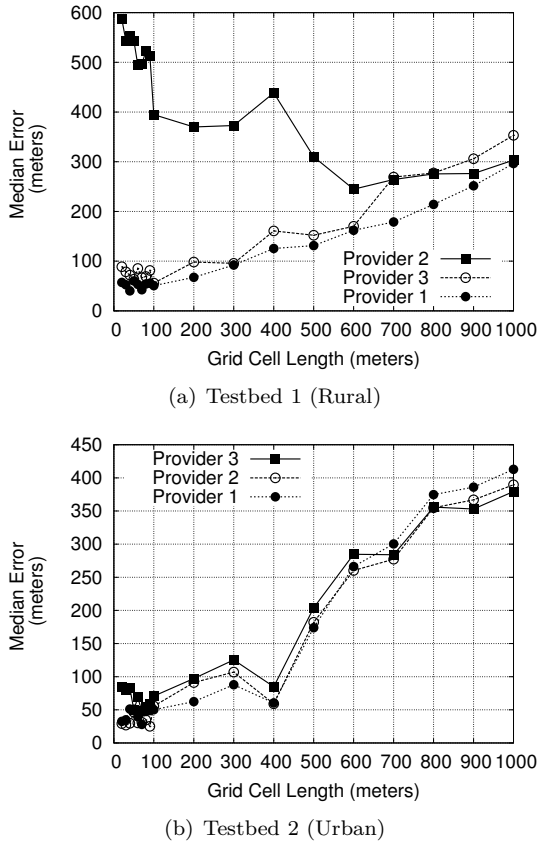

Figure 9. Effect of using different network providers on *CellSense*'s median error.

Provider 2's performance increases with the increase of the grid cell length until it reaches an optimal point at 600m and then decreases again. This is due to two opposing factors: (1) As the grid cell length increases, we have more samples to construct the histogram, leading to better histograms and accuracy. (2) As the grid cell length increases, the fingerprint density decreases and accuracy decreases. This behavior is not noticed with the other providers as they have a higher cell tower density that makes the second factor the dominating factor.

## C. Results for the Hybrid Technique

In this section, we compare the performance of the *CellSense-Hybrid* technique described in Section III-E to the basic *CellSense* technique. The *CellSense-Hybrid* technique mixes both *CellSense* and a deterministic technique in its two phases. Figure 10 shows that the accuracy of *CellSense* degrades as the grid cell length increases since the points inside a cell become further away from its centroid, increasing the estimation error and reducing accuracy. On the other hand, the *CellSense-Hybrid* technique has a robust performance, in terms on accuracy, for different grid sizes under the two testbeds. This is due to the estimation refinement phase.

The figure also shows that the running time of the *CellSense* technique decreases quadratically with the cell size. On the other hand, there are two factors affecting the running time of the *CellSense-Hybrid* technique. (1) As the grid size increases, the number of cells decreases and hence the running time of the first phase of the algorithm decreases. (2) However, as the grid size increases, the number of fingerprint points inside a cell increases and, consequently, the time for the second phase of the algorithm. This leads to the minimum point for the running time at $G = 70$ in the figure.

## D. Comparison with Other Techniques

In this section, we compare the performance of the *CellSense* and the *CellSense-Hybrid* techniques, in terms of running time, localization error, and complexity, to other RSSI-based GSM localization techniques described in Section II-F. Table II summarizes the parameters that achieve the best performance for all techniques. For the percentage enhancement numbers, our reference is the technique that achieves the best value. Therefore, we used *CellSense* as the reference in accuracy and *CellSense-Hybrid* as the reference in running time.

*1) Localization Error:* Figure 11 shows the CDF of distance error for the different algorithms for the two testbeds. Table III summarizes the results. The table shows that our proposed techniques are better than any other technique with at least 108.57% in rural areas and at least 89.30% in urban areas. All techniques perform better in urban areas than rural areas due to the higher density of cell towers and the more differentiation between fingerprint locations due to the dense urban area structures. This is excluding the *CellSense-Hybrid* technique, whose accuracy is consistent between the two testbeds. The loss of accuracy of the *CellSense-Hybrid* technique, as compared to *CellSense*, comes at significant gains in running time as quantified in the next section.

*2) Running time:* Figure 12 compares all algorithms in terms of the average time required for one location estimate. Table III summarizes the results. The results show that the proposed techniques significantly outperform the other techniques by at least 506.21% in rural areas and at least 440.19% in urban areas. All techniques take more time on average in the urban testbed than in rural testbed due to the increase in the number of cell towers. The cell-ID based technique, i.e. Google's MyLocation, has a consistent running time as it depends on the associated cell tower ID only. Although its time involves communicating with Google servers over the network, the average running time is much less than a typical network delay. We believe that this is due to the fact that the Location API on the phone returns a cached location as long as the associated cell tower does not change. The Gaussian processes approach is the most demanding technique in terms of the running time. The *CellSense-Hybrid* technique provides about three to five times enhancement in the running time compared to the *CellSense* technique.

*3) Complexity Analysis:* In this section, we analyze the algorithmic complexity of all techniques. Table III summarizes the results.

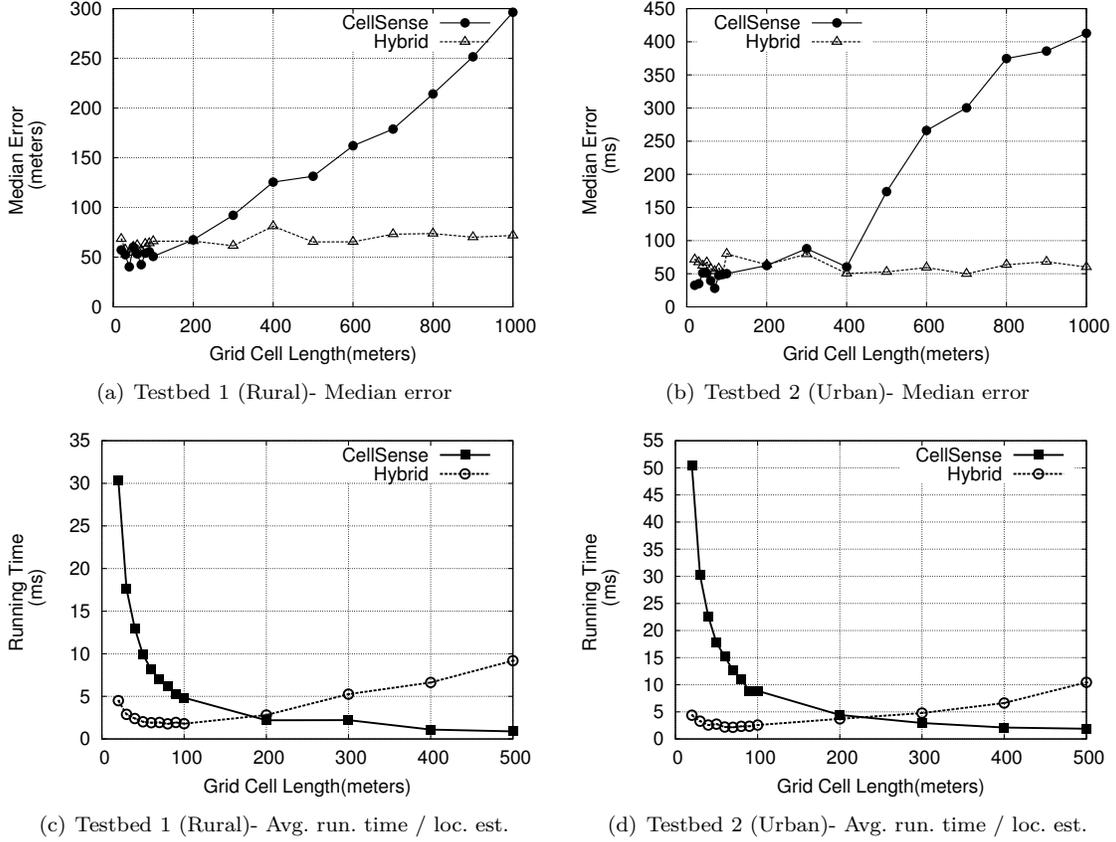

Figure 10. Comparison between *CellSense-Hybrid* and *CellSense* techniques under the two testbeds.

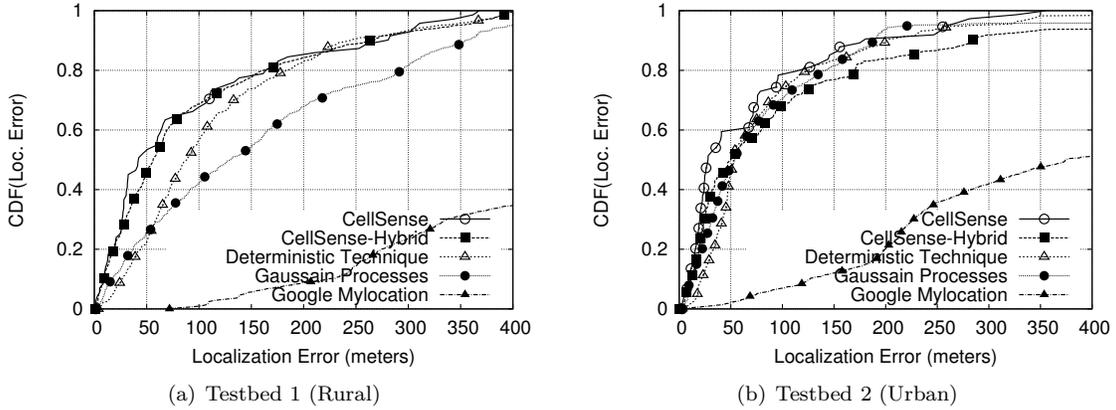

Figure 11. CDF's of distance error for different techniques under the two testbeds. The tails of the CDF's are truncated for clarity of presentation.

- *Google's MyLocation*: is a cell-ID based technique. It has $O(1)$ complexity as it is probably a hash table lookup for the location of the cell tower the phone is connected to. However, we do not have more details from *Google* to confirm our hypothesis.
- *CellSense*: To compute the probability of each grid cell, we need $O(qN_sN_c)$ operations. Computing the weighted average of the most probable $K$ locations, using an order statistics algorithm, requires $O(KN_c)$ for small $K$. Therefore, we need $O((qN_s + K)N_c)$ operations in total for each location estimate.
- *Deterministic technique*: Similar to the *CellSense* technique, it requires $O((qN_s + K)N_c)$.
- *Gaussian processes*: To compute the probability of each precomputed point we need $O(tN_p)$. Computing the weighted average of all the precomputed locations requires $O(N_p)$ operations. Therefore, the overall all algorithm requires $O(tN_p)$ per location estimate.
- *CellSense-Hybrid*: Calculating the probability of each grid cell in the first phase takes $O(qN_c)$. To apply the K-nearest neighbor algorithm in the second phase inside the most probable cell we need $O((q + K)N_0)$.

| Algorithms | Google's MyLocation | Deterministic | Gaussian Processes | *CellSense* | *CellSense-Hybrid* |
|---|---|---|---|---|---|
| Testbed 1 (Rural) Median Error (m) | 656.37 (1446.94%) | 88.50 (108.57%) | 130.50 (207.56%) | 42.43 (**Reference**) | 54.4 (32.33%) |
| 95th percentile (m) | 2767.6 | 328.74 | 401.6 | 308.56 | 354.36 |
| Testbed 2 (Urban) Median Error (m) | 374.58 (1244.51%) | 52.74 (89.30%) | 53.92 (93.54%) | 27.86 (**Reference**) | 56.19 (101.69%) |
| 95th percentile (m) | 3927.8 | 280.06 | 227.68 | 273.14 | 290.45 |
| Testbed 1 (Rural) Avg. time/loc. (ms) | 12.04 (580.22%) | 10.73 (506.21%) | 2873.41 (162240%) | 7.04 (297.74%) | 1.77 (**Reference**) |
| Testbed 2 (Urban) Avg. time/loc. (ms) | 12.04 (462.62%) | 11.56 (440.19%) | 3148.35 (147019%) | 12.66 (491.59%) | 2.14 (**Reference**) |
| Complexity | $O(1)$ | $O((qN_s + K)N_c)$ | $O(qN_p)$ | $O((qN_s + K)N_c)$ | $O(qN_c + (q+K)N_0)$ |

Table III
COMPARISON BETWEEN DIFFERENT TECHNIQUES USING THE TWO TESTBEDS. NUMBERS BETWEEN PARENTHESIS REPRESENT PERCENTAGE DEGRADATION COMPARED TO THE REFERENCE TECHNIQUE(THE BEST TECHNIQUE). $q$ IS THE NUMBER OF CELL TOWERS. $N_s$ IS THE NUMBER OF SUCCESSIVE SAMPLES USED IN ESTIMATION. $N_p$ IS THE NUMBER OF PRECOMPUTED POINTS IN THE GAUSSIAN PROCESSES TECHNIQUE. $N_c$ IS NUMBER OF GRID CELLS. $N_0$ IS NUMBER OF SAMPLES IN THE MOST PROBABLE CELL.

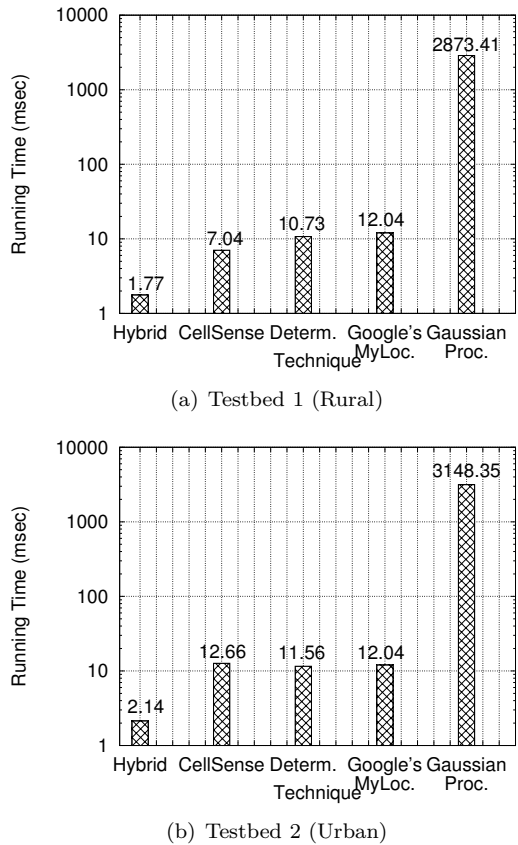

Figure 12. Running time for different techniques under the two testbeds (log scale).

Therefore, we need $O(qN_c + (q+K)N_0)$ operations in total for each location estimate.

Since $N_p$ is typically $>> N_c$ (Table II), to achieve reasonable accuracy, the Gaussian processes approach is very slow compared to the other non-cell-ID based techniques (Figure 12).

Comparing *CellSense* to *CellSense-Hybrid*, we note that for a low grid cell length (the typical operation scenario for *CellSense-Hybrid*), $N_c$ (number of grid cells) is $>> N_o$ (number of samples inside a cell). Therefore, *CellSense-Hybrid* computational overhead is much lower. The opposite is true for high grid cell sizes.

*E. Summary*

In this section, we evaluated the performance of the proposed *CellSense* and *CellSense-Hybrid* techniques. Our results show that the *CellSense-Hybrid* technique has comparable accuracy to the *CellSense* technique with significantly lower computational complexity compared to other techniques.

For the *CellSense* technique, as the grid cell size increases, the performance degrades. Increasing the number of samples used in estimation or the number of averaged most probable locations have a positive effect on accuracy. Increasing the cell towers density has a more positive effect on accuracy than increasing the density of the fingerprint. The good news is that, even though we do not have control on the cell towers density, reducing the fingerprint density by up to 60% still gives good accuracy.

The performance of the *CellSense-Hybrid* technique is consistent over different grid sizes and testbeds. This is due to the estimation refinement phase.

The accuracy of the localization technique under a certain cellular provider is correlated with the provider's cell tower density.

Typically, there is always a trade off between computational overhead and accuracy. However, the *CellSense-Hybrid* techniques provides a good balance between both accuracy and complexity. Its high accuracy comes from its ability of returning one of the original fingerprint points, rather than the center of mass of all locations inside the most probable cell. Its computational advantage at small grid sizes, compared to the *CellSense* technique, comes from using only one sample in the first phase, as compared to $N_s$ samples.

V. CONCLUSION

We proposed *CellSense*, a probabilistic RSSI-based fingerprinting approach for GSM cell phones localization.

We presented the details of the system and how it constructs the probabilistic fingerprint without incurring any additional overhead. We also proposed a hybrid approach that combines probabilistic and deterministic techniques to achieve both high accuracy and low computational requirements.

We implemented our system on Android-based phones and compared it to other GSM-localization systems under two different testbeds. Our results show that the *CellSense-Hybrid* technique's accuracy is better than other techniques with at least 108.57% in rural areas and at least 89.03% in urban areas with more than 5.4 times saving in running time compared to the state of the art RSSI-based GSM localization techniques . We also studied the effect of different parameters on the performance of the system and how the cell towers density and fingerprint density affect accuracy.

Currently, we are working on extending our system in different directions including using parametric distributions, clustering of fingerprint locations, experimenting with larger datasets, comparison with other city-wide commercial systems, targetting low-end phones [19], among others.

## References

[1] M. Ibrahim and M. Youssef, "CellSense: A probabilistic RSSI-based GSM positioning system," in *GLOBECOM*, 2010.
[2] P. Enge and P. Misra, "Special issue on GPS: The Global Positioning System," *Proceedings of the IEEE*, pp. 3–172, January 1999.
[3] S. Tekinay, "Special issue on Wireless Geolocation Systems and Services," *IEEE Communications Magazine*, April 1998.
[4] Y.-C. Cheng, Y. Chawathe, A. LaMarca, and J. Krumm, "Accuracy characterization for metropolitan-scale wi-fi localization," in *MobiSys '05: Proceedings of the 3rd international conference on Mobile systems, applications, and services*. New York, NY, USA: ACM, 2005, pp. 233–245.
[5] I. Smith, J. Tabert, A. Lamarca, Y. Chawathe, S. Consolvo, J. Hightower, J. Scott, T. Sohn, J. Howard, J. Hughes, F. Potter, P. Powledge, G. Borriello, and B. Schilit, "Place lab: Device positioning using radio beacons in the wild," in *Proceedings of the Third International Conference on Pervasive Computing*. Springer, 2005, pp. 116–133.
[6] Skyhook wireless, "http://www.skyhookwireless.com."
[7] R. R. C. Ionut Constandache and I. Rhee, "Towards mobile phone localization without war-driving," in *IEEE Infocom*, 2010.
[8] R. S. Andrew Offstad, Emmett Nicholas and R. R. Choudhury, "Aampl: Accelerometer augmented mobile phone localization," in *ACM MELT Workshop (with Mobicom 2008)*, 2008.
[9] I. C. Martin Azizyan and R. R. Choudhury, "Surroundsense: Mobile phone localization via ambience fingerprinting," in *ACM MobiCom*, 2009.
[10] Wikipedia, "Comparison of mobile phone standards — Wikipedia, the free encyclopedia," 2010, [Online; accessed 25-March-2010]. [Online]. Available: \url{http://en.wikipedia.org/wiki/Comparison_of_mobile_phone_standards}
[11] M. Y. Chen, T. Sohn, D. Chmelev, D. Haehnel, J. Hightower, J. Hughes, A. Lamarca, F. Potter, I. Smith, and A. Varshavsky, "Practical metropolitan-scale positioning for GSM phones," in *Proceedings of the Eighth International Conference on Ubiquitous Computing (UbiComp*. Springer, 2006, pp. 225–242.
[12] A. Varshavsky, M. Y. Chen, E. de Lara, J. Froehlich, D. Haehnel, J. Hightower, A. LaMarca, F. Potter, T. Sohn, K. Tang, and I. Smith, "Are GSM phones THE solution for localization?" in *WMCSA '06: Proceedings of the Seventh IEEE Workshop on Mobile Computing Systems & Applications*. Washington, DC, USA: IEEE Computer Society, 2006, pp. 20–28.
[13] E. Elnahrawy, J. Austen-francisco, and R. P. Martin, "Adding angle of arrival modality to basic RSS location management techniques," in *In Proceedings of IEEE International Symposium on Wireless Pervasive Computing (ISWPCS07)*, 2007.
[14] E. Elnahrawy, J. austen Francisco, and R. P. Martin, "Poster abstract: Bayesian localization in wireless networks using angle of arrival," in *Proceedings of the Third ACM Conference on Embedded Networked Sensor Systems (SenSys'05)*, 2005.
[15] P. Biswas, H. Aghajan, and Y. Ye, "Integration of angle of arrival information for multimodal sensor network localization using semidefinite programming," in *In Proceedings of 39th Asilomar Conference on Signals, Systems and Computers*, 2005.
[16] M. Li and Y. Lu, "Angle-of-arrival estimation for localization and communication in wireless networks," 2008.
[17] Google Maps for Mobile, "http://www.google.com/mobile/maps/."
[18] M. Youssef, M. A. Yosef, and M. N. El-Derini, "GAC: Energy-efficient hybrid gps-accelerometer-compass gsm localization," in *GLOBECOM*, 2010.
[19] M. Ibrahim and M. Youssef, "A hidden markov model for localization using low-end GSM cell phones," in *ICC*, 2011.

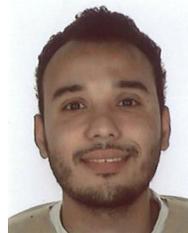

**Mohamed Ibrahim** received his B.Sc. in computer science from Alexandria University, Egypt in 2009 and a M.Sc. in wireless technology from Nile University, Egypt in 2011. He is now a PhD candidate in University of technology of Troyes, France. His research interests include location determination technologies, sensor networks, and pattern recognition.

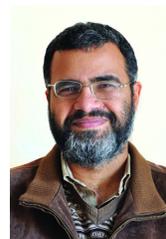

**Moustafa Youssef** is an Assistant Professor at Alexandria University and Egypt-Japan University of Science and Technology (E-JUST), Egypt. He received his Ph.D. degree in computer science from University of Maryland, USA in 2004 and a B.Sc. and M.Sc. in computer science from Alexandria University, Egypt in 1997 and 1999 respectively. His research interests include location determination technologies, pervasive computing, sensor networks, and network security. He has eight issued and pending patents. He is an area editor of the ACM MC2R and served on the organizing and technical committees of numerous conferences and published over 70 technical papers in refereed conferences and journals. Dr. Youssef is the recipient of the 2003 University of Maryland Invention of the Year award for his *Horus* location determination technology and the 2010 TWAS-AAS-Microsoft Award for Young Scientists, among others.